\documentclass[conference,romanappendices,10pt]{IEEEtran}
\IEEEoverridecommandlockouts

\usepackage{algorithmic,amsmath,amssymb,amsthm,array,bbm,bibentry,cite,color,comment,dsfont}
\usepackage{enumerate,enumitem,eurosym,float,graphicx,lettrine,mathrsfs,multirow,nomencl}
\usepackage{pict2e,psfrag,ragged2e,relsize,setspace,subfigure,supertabular,tabularx,url}
\usepackage[english]{babel}

{		}
{ 		}
{ 		}
{ 		\newtheorem{proposition}{Proposition}}
{ 		}
{		}
{		}
{ 		}

\renewcommand{\d}{\mathrm{d}}

\newcommand{\s}{\mathbf{s}}

\newcommand{\E}{\mathbf{E}}

\renewcommand{\P}{\mathbf{P}}

\newcommand{\setA}{\mathcal{A}}

\newcommand{\setM}{\mathcal{M}}

\newcommand{\setS}{\mathcal{S}}

\newcommand{\ex}{\operatornamewithlimits{\mathsf{E}}}

\newcommand{\sir}{\mathrm{SIR}}

\newcommand{\nop}[1]{}

\def\Ps{P_{\rm s}}
\def\P{\mathbb{P}}
\def\E{\mathbb{E}}

\begin{document}

\title{QoS Provisioning in Large Wireless Networks}

\author{\IEEEauthorblockN{Marios~Kountouris\IEEEauthorrefmark{1}, Nikolaos~Pappas\IEEEauthorrefmark{2}, and Apostolos~Avranas\IEEEauthorrefmark{1}} %
	
	\IEEEauthorblockA{\IEEEauthorrefmark{1}Mathematical and Algorithmic Sciences Lab, Paris Research Center, Huawei France}
	\IEEEauthorblockA{\IEEEauthorrefmark{2}Department of Science and Technology, Linkoping University, Sweden} %
	
	Email: \{marios.kountouris,apostolos.avranas\}@huawei.com, nikolaos.pappas@liu.se
}

\maketitle

\begin{abstract}
Quality of service (QoS) provisioning in next-generation mobile communications systems entails a deep understanding of the delay performance. The delay in wireless networks is strongly affected by the traffic arrival process and the service process, which in turn depends on the medium access protocol and the signal-to-interference-plus-noise ratio (SINR) distribution. In this work, we characterize the conditional distribution of the service process given the point process in Poisson bipolar networks. We then provide an upper bound on the delay violation probability combining tools from stochastic network calculus and stochastic geometry. Furthermore, we analyze the delay performance under statistical queueing constraints using the effective capacity formulation. The impact of QoS requirements, network geometry and link distance on the delay performance is identified. Our results provide useful insights for guaranteeing stringent delay requirements in large wireless networks.
\end{abstract}

\vspace{1mm}


\section{Introduction} \label{sec:intro}
Data traffic has been growing tremendously over the last decade, fueled by the ubiquity of smart devices and bandwidth-demanding applications. Current wireless networks are confronted with an avalanche of heterogeneous traffic with diverse requirements in terms of rate, reliability and latency. Emerging mobile communication systems will not only be designed to provide enhanced spectral efficiency and coverage, but they should also meet the delay requirements of new delay-sensitive applications, such as industrial control, automated driving and healthcare. 
Different applications are expected to have very diverse QoS requirements in terms of throughput and delay. QoS provisioning is instrumental for next-generation low latency networks; yet it is particularly challenging mainly due to time-varying wireless channels and spatio-temporal randomness in traffic arrivals and interferers’ locations. Ensuring deterministic (hard) QoS guarantees would most likely result in extremely conservative performance. As a result, providing statistical (soft) QoS guarantees, in terms of effective bandwidth/capacity and bounds in queue length and delay violation probability, stands as a powerful approach to characterize delay QoS provisioning in wireless networks \cite{Chang94_AutControl,Chang00,WuNegi03_EC,Tang08}.

There have been several attempts at quantifying delay in wireless networks, and queueing theory has been instrumental in providing exact backlog and delay characterization. However, queueing analysis is largely restricted to networks with a single or few interacting queues with random arrivals and typically provides the average delay rather than the worst-case delay, which is of cardinal importance in mission-critical applications.  
The delay in wireless networks is strongly affected by the queueing process and the service process of the packets. The latter is mainly governed by the access protocol and the link quality, i.e., the received SINR. The SINR in turn critically depends on the link distance and the network geometry on which the path loss and the fading characteristics are dependent upon. 
Although the locations of nodes in wireless networks are traditionally modeled by regular grids, the spatial node distribution in emerging networks (e.g. heterogeneous cellular networks) is irregular. Stochastic geometry and point process theory have recently proved to be powerful mathematical tools for analyzing and designing large wireless networks with spatial randomness. These approaches have focused on metrics such as coverage probability and spatial average rate, and calculate spatial averages by considering a snapshot of the network. Analyzing the delay in large spatial networks is very challenging due to the correlation of interference \cite{Radha_Corr,Radha_linkoutage,IntfCorrelation} and the large number of interacting queues. The local delay, i.e., the random number of transmission attempts until a packet is successfully transmitted to its target receiver, is proposed in \cite{BB_PhaseTrans} and extended in \cite{Haenggi_LD}. Local delay assumes fully backlogged nodes, thus it captures the transmission delay, but not the queueing delay. Throughput maximization subject to delay constraints in two-tier spatial networks is studied in \cite{ZhengWoWMoM}.

In this paper, we investigate the delay performance of large wireless networks in the presence of statistical QoS constraints, which are imposed as limits on the delay violation and buffer overflow probabilities. We start by characterizing the distribution of the service process in Poisson bipolar networks. The results are exploited in order to derive a bound on the delay violation probability using tools from stochastic network calculus \cite{Chang00,Jiang08book,Fidler15_Guide} and the effective capacity. For relevant results, we perform space-time scale separation and consider static or low-mobility spatial networks. This results in temporal interference correlation and the conditional rate given the point process will vary from node to node. Thus, both delay violation probability and effective capacity are random variables whose statistics ought to be found. Our analytical results show the effect of inter-node distances, spatial randomness, and QoS constraints on the delay violation and effective capacity performance of large spatial networks.

\section{System Model} \label{sec:syst}
\subsection{Network Model}
We consider a communication network in which the locations of the (potential) transmitters are modeled as a homogeneous Poisson point process (PPP) $\Phi = \{x_i\} \subset \mathbb{R}^2$ of intensity $\lambda$. Each transmitter has an associated receiver at fixed distance $r$ in a random direction (denoted by $R_x$ for a transmitter $x$). This model is commonly referred to as Poisson bipolar network.

The small scale fading between two nodes is independent and identically distributed (i.i.d.) across time and space (unless otherwise stated) and is exponentially distributed (Rayleigh fading). The transmit power at all nodes is fixed to 1.
The large-scale path loss function is denoted by $\ell(x):\mathbb{R}^2 \to [0,\infty]$ and is assumed to be a non-increasing function of $\| x \|$ and $\int_{B(o,d)}\ell(x)\d x < \infty$, $\forall d$, where $B(o,d)$ is the ball of radius $d$ centered around the origin $o = (0,0)$. In this paper, we focus on a non-bounded model $\ell(x) = \| x \|^{-\alpha}$, $\alpha > 2$. 
Time is divided into discrete slots of equal duration and transmission attempts are synchronized. We focus on the interference-limited case, but our analysis can be easily extended including background noise. 

The received signal-to-interference ratio (SIR) in time slot $t$ is given by
\begin{eqnarray} \label{eq:SIR}
\sir_{R_x,t} = \frac{h_{xR_x}^{t} \ell(r)}{\sum_{y \in \Phi_{t} \setminus \{x\}} h_{yR_x}^{t}\ell(y-R_x)}
\end{eqnarray}
where $h_{xy}^t$ is the small-scale fading coefficient between nodes $x$ and $y$ in time slot $t$, and $\Phi_t\subset \Phi$ is the set of active interferers in time slot $t$. The interference at time slot $t$ can be alternatively written as 
\begin{eqnarray} \label{eq:It}
I_{R_x,t} = \sum_{y \in \Phi \setminus \{x\}} h_{yR_x}^t\ell(y-R_x)\mathds{1}(y \in \Phi_t)
\end{eqnarray}
where $\mathds{1}(\cdot)$ is the indicator function.
 
The success probability of a typical link is given by 
\begin{equation}
\P^{!o}(\sir_{R_o,t} > \xi) = \lim_{\delta\to\infty}\frac{\displaystyle\sum_{x\in \Phi \cap B(o,\delta)}\P(\sir_{R_x,t} > \xi \mid \Phi)}{\lambda\pi p \delta^2}.
\label{Ps}
\end{equation}
We focus on the typical link with a transmitter located at the origin and we drop the time and node subindexes whenever evident.

We denote $\Ps(\xi)\triangleq \P(\sir > \xi \mid \Phi)$ the success probability given the point process (i.e., conditioned on the location of interferers) and that the transmitter of interest is active, which is taken over the fading and the random access scheme. The conditional probability $\Ps(\xi)$ can be interpreted as the mark of a virtual typical link placed at the origin, whose distribution is given by $\P^{!o}(\Ps(\xi) > x)$, $x\in [0, 1]$, where $\P^{!o}$ is the reduced Palm probability (for a PPP we have $\P^{!o} = \P$) \cite{DaleyVereJones}. 

\subsection{Traffic Model}
We consider a system-theoretic stochastic model as in \cite{Multihop13_Infocom}, which involves a queueing system with stochastic arrival and departure processes described by bivariate stochastic processes $A(\tau, t)$ and $D(\tau, t)$, respectively. A fluid-flow traffic model is adopted and the system starts with empty queues at $t=0$.
The number of bits arriving at the queue at a discrete time instant $i$ is modeled by the arrival process $a_i$. For successful transmissions, the service process is equal to the instantaneous capacity. In case of transmission errors, the service is considered to be zero as no data is removed from the queue. The departure process $d_i$ describes the number of bits that arrive successfully at the destination and depends on both the service process and the number of bits waiting in the queue. Acknowledgments and feedback messages are assumed to be instantaneous and error-free.  
To avoid data loss, data is stored in a buffer or queue, in which it will remain for a random time until the receiver indicates that data was successfully decoded. 
At each time slot, node $j \in \Phi$ independently transmits with probability $p_j$. The steady-state probability $p_j$, $\forall j \in \Phi$ depends on the arrival process of $j$-th node, the transmit probability of the other nodes $p_i$, $\forall i \in \Phi, i \neq j$, and the channel of all nodes. A node remains idle when there is no traffic arrival and the queue is empty due to the early arrival and late departure assumption. The aforementioned two events can be assumed independent under non-saturated or light traffic conditions. Unless otherwise stated, we set $p_i = p$, $\forall i \in \Phi \setminus\{o\}$.

The cumulative arrival and departure processes for any $0 \leq \tau \leq t$, measured in bits of the flow during time interval $[\tau,t)$, are defined as
\begin{eqnarray}
A(\tau, t) = \displaystyle \sum_{i = \tau}^{t-1}a_i, \ \ \text{and} \ \ D(\tau, t) = \displaystyle \sum_{i = \tau}^{t-1}d_i.
\end{eqnarray} 

For lossless first-in first-out (FIFO) queueing systems, the delay $W(t)$ at time $t$, i.e., the number of slots it takes for an information bit arriving at time $t$ to be received at the destination, is defined as 
\begin{eqnarray}
W(t) = \inf\{u > 0 : A(0,t)/D(0,t+u) \leq 1 \}.
\end{eqnarray} 
and the delay violation probability is given by 
\begin{eqnarray*}
\Lambda(w,t) = \displaystyle \sup_{t\geq 0}\mathbb{P}\left[W(t) > w \right].
\end{eqnarray*}

\section{Service Process Characterization} \label{sec:perf}
The instantaneous rate or capacity $C_t$ of the channel at time $t$ can be expressed as a function of the instantaneous SNR or SIR at this time. 
Assuming flat-fading, Gaussian codebooks and ideal link adaptation, the instantaneous rate can be expressed as
\begin{eqnarray}
C_t=N\log(1+\sir_t)  \ \ \  (\rm {nats/s})
\end{eqnarray}
where $N$ is the number of transmitted symbols per time slot. The symbol rate is usually related to the bandwidth $W$ as $N=2W$ (Shannon-Harltley theorem). In the remainder, we assume $N=1$ to simplify notation and we reincorporate this parameter into
the equations in the numerical results. 

The service process (or cumulative capacity) through period $(\tau,t]$ is defined as 
\begin{eqnarray}
S(\tau,t) \triangleq \sum\limits_{i=\tau}\limits^{t-1}{C_i},
\end{eqnarray} 
and is a random variable with cumulative distribution function (cdf) $F_{S(\tau,t)}(x) = \P(S(\tau,t) \leq x)$, $x>0$.

If $C_i$ and $C_j$, $i\neq j$, are independent, then $f_{S(\tau,t)} = f_{C_{\tau+1}} \ast \ldots \ast f_{C_t}$, where $\ast$ denotes the convolution operation, i.e., $(f\ast g)(x) = \int_{-\infty}^{+\infty}f(x-y)g(y)\d y$.
Hence, 
\begin{equation}
F_{S(\tau,t)}(x) = \int_{-\infty}^{x} f_{S(\tau,t)}(y) \d y.
\end{equation}
An upper bound can be derived using Young's inequality, which states $\| f*g \|_r \leq \| f \|_p \| g \|_q$ for $1/p+1/q=1/r+1$, $f\in L^p(\mathbb{R}^d)$ and $g\in L^q(\mathbb{R}^d)$. 
When all marginal distributions are identical ($F_{C_i} \sim F_C, \forall i$), the probability density function (pdf) of the service process is given by the $n$-fold convolution with $n=t-\tau$, i.e., $f_{S(\tau,t)} = f_{C}^{\ast n}$.

When the length of the period $t-\tau$ is large, $F_{S(\tau,t)}(x)$ converges to a normal distribution (Central Limit Theorem)
\begin{eqnarray}\label{tccdf}
F_{S(\tau,t)}(x) \approx Q\left(\frac{x-\E[S(\tau,t)]}{\sigma^{2}[S(\tau,t)]}\right), 
\end{eqnarray}
with mean $\E[S(\tau,t)]= \sum\limits_{i=\tau+1}\limits^{t}\E[C_i]$ and variance $\sigma^{2}[S(\tau,t)]= \sum\limits_{i=\tau+1}\limits^{t}\sigma^{2}[C_i]$, where $Q(x) \triangleq \int_{-\infty}^{x}\frac{1}{\sqrt{2\pi}}e^{-y^2/2}\d y$.

If $C_i$ and $C_j$, $i\neq j$, are not independent, the exact calculation of the service rate distribution seems to be highly involved. For that, we resort to the Fr\'{e}chet-Hoeffding bounds on copulas \cite{CopulasBook,CumulativeCop}, which give that cdf of the cumulative capacity satisfies 
\begin{equation}
F_{S(\tau,t)}^{l}(z)\le F_{S(\tau,t)}(z)\le F_{S(\tau,t)}^{u}(z) \label{dcdf}
\end{equation}
where 
\begin{eqnarray*}
F_{S(\tau,t)}^{u}(z) &\triangleq& \inf_{\sum\limits_{i=\tau+1}\limits^{t}z_{i}=z}\left[
\sum\limits_{i=\tau+1}\limits^{t}{F_{C_i}(z_{i})}\right]_{1},
\label{dtccdf}\\
F_{S(\tau,t)}^{l}(z) &\triangleq& \sup_{\sum\limits_{i=\tau+1}\limits^{t}z_{i}=z}\left[
\sum\limits_{i=\tau+1}\limits^{t}{F_{C_i}(z_{i})}-(t-\tau-1) \right]^{+}. \label{dtccdfl}
\end{eqnarray*}
where $[f]^+ = \max(f,0)$, $[f]_1 = \min(f,1)$, and $F_{C_i}(z) = \P(\sir_i \leq e^z-1) = F_{\sir_i}(e^z-1) = 1 - \Ps(e^z-1)$, $z > 0$.

\subsection{Moment Generating Function}
The service rate distribution can also be characterized via the moment generating function (MGF) $M_{S(\tau,t)}(\theta)$, $\theta \in \mathbb{R}$, which is given by
\begin{eqnarray*}
	M_{S(\tau,t)}(\theta)&\triangleq&{\E\left[e^{\theta{S(\tau,t)}}\right]} 
	= \int_{-\infty}^{\infty} {e^{\theta{z}}} \d{F_{S(\tau,t)}(z)}.
	\end{eqnarray*}
For the independent case, we have $\overline{M}_{S(\tau,t)}^{\rm ind}(\theta) = \prod_{i=\tau}^{t-1}\overline{M}_{C_i}(\theta)$ and for the i.i.d. case, we have $\overline{M}_{S(\tau,t)}^{\rm iid}(\theta) = (\overline{M}_{C_i}(\theta) )^{t-\tau}$, where $\overline{M}_{S(\tau,t)}(\theta)=M_{S(\tau,t)}(-\theta)$.

The distribution of the service process can be calculated using Gil-Pelaez theorem \cite{GilPelaez}
\begin{eqnarray*}
F_{S(\tau,t)}(z)  & = & \frac{1}{2}-\frac{1}{\pi}\int_{0}^{\infty}\Im[e^{-jtx}M_{S(\tau,t)}(jt)]\frac{\d t}{t} \\
& = & \frac{1}{2\pi j}\int_{c-j\infty}^{c+j\infty}e^{\log M_{S(\tau,t)}(t)-tx}\frac{\d t}{t}
\end{eqnarray*}
with $c \in \mathbb{R}_{> 0}$ in the convergence strip of the cumulant generating function $\log M_{S(\tau,t)}(t)$.
The distribution can be conveniently evaluated numerically using L\'{e}vy's inversion theorem \cite{Feller_vol2}.

\subsection{Mellin transform}
For calculating the delay violation probability using stochastic network calculus, the service process is characterized in terms of its Mellin transform (MT). 

The MT of a nonnegative random variable $X$ is defined as $\mathcal{M}_X(s) \triangleq \E\left[X^{s-1}\right] = \overline{M}_{\log X}(s-1)$ for any $s \in \mathbb{C}$ for which the expectation exists. For the service process, we have 
\begin{eqnarray*}
	{\mathcal{M}_{\mathcal{S}(\tau,t)}(s) \triangleq \E[({\mathcal{S}(\tau,t))^{s-1}}]} = \int_{-\infty}^{\infty} {z^{s-1}} \d F_{\mathcal{S}(\tau,t)}(z). 
\end{eqnarray*}

According to (\ref{tccdf}) and (\ref{dcdf}), the MT of the service process for the independent and dependent case is given by, respectively
\begin{IEEEeqnarray*}{rCl}
	\IEEEeqnarraymulticol{3}{l}
	{\mathcal{M}_{\mathcal{S}(\tau,t)}^{\rm ind}(s) \approx \int_{-\infty}^{\infty} { z^{s-1} } \d Q\left(\frac{z-\E[S(\tau,t)]}{\sigma^{2}[S(\tau,t)]}\right),} \label{mellini} \\
	\IEEEeqnarraymulticol{3}{l}
	{\mathcal{M}_{\mathcal{S}(\tau,t)}^{{\rm dep},l}(s)=\int_{-\infty}^{\infty} {  z^{s-1}} \d F_{S(\tau,t)}^{l}(z)\le \mathcal{M}_{\mathcal{S}(\tau,t)}^{\rm dep}(s)} \label{mellindl}\\\qquad\qquad\quad
	\leq \int_{-\infty}^{\infty} { z^{s-1} } \d F_{S(\tau,t)}^{u}(z)= \mathcal{M}_{\mathcal{S}(\tau,t)}^{{\rm dep}, u}(s) \label{mellindu}
\end{IEEEeqnarray*}
where the upper and lower bounds hold for $s<1$.

\section{Delay Performance in Static Networks} \label{sec:delay}
In this section, we analyze the QoS performance in terms of delay violation probability and effective capacity. We consider a static network, where the random locations of nodes do not vary with time, and perform the analysis given the locations of the nodes, i.e., conditioned on $\Phi$. 
The conditional success probability $\Ps(\xi)$ is a random variable that depends on the spatial distribution, which implies that the conditional service rate varies from node to node. This means that some nodes have an arbitrarily small service rate and consequently an arbitrarily large delay. Instead of deriving spatial averages for the delay metrics, we aim at obtaining the spatial distribution of the delay violation probability and of the effective capacity. For that, we derive the conditional service rate and calculate the delay metrics for each draw of points in the space. As a result, the delay violation probability and the effective capacity are random variables and we are interested in deriving their distribution.

\subsection{Delay Violation Probability}
In this section, we obtain an upper bound on the delay violation probability using a statistical characterization of the arrival and service processes in the exponential (or SIR) domain \cite{Multihop13_Infocom}. First, we convert the cumulative processes in the bit domain to the SIR domain (denoted by calligraphic letters) through the exponential function, i.e., 
\begin{equation*}
\mathcal{A}(\tau, t) = e^{A(\tau, t)}, \quad \mathcal{D}(\tau, t) = e^{D(\tau, t)}, \quad \mathcal{S}(\tau, t) = e^{S(\tau, t)}.
\end{equation*}

An upper bound on the delay violation probability can be computed by means of the Mellin transforms of $\mathcal{A}(\tau, t)$ and $\mathcal{S}(\tau, t)$ \cite{Multihop13_Infocom}:
\begin{equation}\label{eq:delay_bound_1}
p_\mathrm{v}(w)  = \inf_{s>0}\left\lbrace K(s,-w)\right\rbrace \geq \Lambda(w)
\end{equation}
where $K(s,-w)$ is the steady-state kernel, defined as
\begin{equation}\label{eq:ker_limits}
\mathcal{K}(s,-w) = \lim_{t\to\infty} \sum_{u=0}^{t}\mathcal{M}_{\mathcal{A}}(1+s,u,t)\mathcal{M}_{\mathcal{S}}(1-s,u,t+w).
\end{equation}

For the arrival process, assuming that $\mathcal{A}(\tau, t)$ has stationary and independent increments, the MT becomes independent of the time instance, i.e.,
\begin{eqnarray*}
\setM_\setA (s,\tau,t) &=& \ex\left[\left(\prod_{i=\tau+1}^{t} e^{a_i}\right)^{s-1}\right] \\
&=& \ex\left[e^{a(s-1)}\right]^{t-\tau} = \setM_\alpha(s)^{t-\tau}
\end{eqnarray*}
where $\alpha = e^a$ denotes the non-cumulative arrival process in the SIR domain.
Using Chang's traffic characterization \cite{Chang00}, we consider the traffic class of $(\sigma(s), \rho(s))$-bounded arrivals, whose MGF in the bit domain is bounded by 
\begin{eqnarray}
\frac{1}{s}\log\E[e^{sA(\tau,t)}] \leq \rho(s)\cdot(t-\tau) + \sigma(s) 
\end{eqnarray}
for some $s>0$. Restricting ourselves to the case where $\rho$ is independent of $s$ and $\sigma(s) = 0$, we have
\begin{equation}\label{eq:defmellin_alpha}
\setM_\alpha(s) = e^{\rho(s-1)}.
\end{equation}

For the service process, as said before, we consider a static network and condition on $\Phi$ (random but static over time). Therefore, the SIRs are conditionally independent and the random variations come from independent block fading for all active links. The MT of the (conditional) service process is given by 
\begin{eqnarray}\label{eq:defmellin_s}
\setM_\setS (s,\tau,t) &=& \E\left[\left(\prod_{i=\tau}^{t-1}(1+\sir_i)\right)^{s-1}\mid \Phi\right] \nonumber \\
&\hspace{-20mm} =& \hspace{-10mm} \E\left[(1+\sir)^{s-1}\mid \Phi\right]^{t-\tau} = \left(\setM_{\gamma}(s)\right)^{t-\tau}.
\end{eqnarray}

Plugging (\ref{eq:defmellin_alpha}) and (\ref{eq:defmellin_s}) into (\ref{eq:ker_limits}) and following \cite{Multihop13_Infocom}, the steady-state kernel can be finally rewritten as 
\begin{eqnarray}
\mathcal{K}(s,-w) = \frac{\left(\mathcal{M}_{\gamma}(1-s)\right)^{w}}{1 - \mathcal{M}_{\alpha}(1+s)\mathcal{M}_{\gamma}(1-s)},
\end{eqnarray} 
for any $s > 0$ under the stability condition $\mathcal{M}_{\alpha}(1+s)\mathcal{M}_{\gamma}(1-s) < 1$. The upper bound on the conditional delay violation probability (\ref{eq:delay_bound_1}) is thus reduced to
\begin{equation}
p_\mathrm{v}(w) = \inf_{s>0}\left\lbrace\frac{\left(\mathcal{M}_{\gamma}(1-s)\right)^{w}}{1 - \mathcal{M}_{\alpha}(1+s)\mathcal{M}_{\gamma}(1-s)}\right\rbrace.
\end{equation}

The above delay bound can be calculated using the following result.  
\begin{proposition} \label{PropMT}
	The MT of the service process in static networks is given by $\mathcal{M}_{S(\tau,t)}(s) = (\mathcal{M}_{\gamma}(\s) )^{t-\tau}$ where 
	\begin{eqnarray*}
	\mathcal{M}_{\gamma}(s) = 	1 +(s-1) \int_0^{\infty}\Ps(y)(1+y)^{s-2}\d y, \ \textrm{for} \  s <1
	\end{eqnarray*}	
	where $\Ps(\xi) = \displaystyle \prod_{x\in\Phi\backslash\{o\}} \Big(\frac{p}{1+\xi r^\alpha\|x-R_o\|^{-\alpha}}+1-p\Big)$.
\end{proposition}

\begin{IEEEproof}
	See Appendix~\ref{sec:A1}.
\end{IEEEproof} \vspace{1mm}

The above semi-closed form expression requires numerical integration. For easier numerical evaluation and in order to gain insights, we provide the following upper bound on $\mathcal{M}_{\gamma}(s)$, which in turn provides an upper bound on the delay violation probability. Taking into account only the interference from the nearest interfering transmitter  $x_{\min} = \arg\min_{x\in \Phi\setminus\{o\}}\|x-R_o\|$, we have 
\begin{eqnarray*}
	\mathcal{M}_{\gamma}^{u1}(s) & \leq & 1 +(s-1) \int_0^{\infty}\frac{(1+y)^{s-2}}{1+y r^\alpha\|x_{\min}-R_o\|^{-\alpha}}\d y \\
	& = & 1 +\frac{(s-1){}_2 F_1\left(2-s,2-s;1-s;1-Z\right)}{Z^{s-1}(2-s)} \\
		& \stackrel{(a)}{\leq} & 1 +(s-1)[(3-2s)Z]^{-1/2} = \mathcal{M}_{\gamma}^{u2}(s)
\end{eqnarray*}
where ${}_2 F_1(a,b;c;x)$ is the Gauss hypergeometric function, $Z = r^\alpha\|x_{\min}-R_o\|^{-\alpha}$ and (a) follows applying Cauchy-Schwarz inequality. 

Calculating the the distribution of the (conditional) delay violation probability $\P^{!o}(p_\mathrm{v}(w) > x)$, $x\in [0, 1]$ is complex. The distribution of the upper bound based on nearest neighbor can be calculated as follows. Since the delay violation probability $p_\mathrm{v}(w)$ is a decreasing function (say $g$) of the random variable $\|x_{\min}\|$, we have  
\begin{eqnarray*}
\P^{!o}(p_\mathrm{v}(w) > x) & \leq & \P^{!o}(p_\mathrm{v}^u(w) > x) = \P^{!o}(g(\|x_{\min}\|) > x) \\
& = &\P^{!o}(\|x_{\min}\| > g^{-1}(x)) = e^{-\lambda \pi (g^{-1}(x))^2}.
\end{eqnarray*} 

\subsection{Effective Capacity}
The effective capacity is defined as the maximum constant arrival rate at a buffer that can be supported by the service process while satisfying statistical QoS requirements specified by the QoS exponent $\theta$ \cite{WuNegi03_EC}.
For time-varying arrival rates, effective capacity specifies the effective bandwidth of the arrival process that can be supported by the channel.

Let $Q$ be the the stationary queue length, then $\theta$ is the decay rate of the tail distribution of the queue length $Q$
\begin{eqnarray}
\lim_{q\to \infty} \frac{\log\P(Q \geq q)}{q} = -\theta
\end{eqnarray}
and from G{\"a}rtner-Ellis Theorem the buffer violation probability for large $q_{\max}$ is approximated as $\P(Q \geq q_{\max}) \approx e^{-\theta q_{\max}}$. Therefore, larger $\theta$ corresponds to more strict QoS constraints, while smaller $\theta$ implies looser constraints.

In block fading channels with coherence time $T$, the effective capacity simplifies to
\begin{eqnarray*}
\mathcal{R}(\theta T) & \triangleq & -\frac{1}{\theta T}\log\E\left[e^{-\theta TC_t}\right] = -\frac{1}{\theta T}\log \setM_{\gamma}(1-\theta T) \\
& = & -\frac{1}{\theta T}\log\E\left[(1+\sir_t)^{-\theta T}\right]
\end{eqnarray*}

Using that for a positive random variable $X$, $\E[X] = \int_{0}^{\infty}\P(X \geq t)\d t$, we can have the following alternative expression for $\Psi(\theta T) = \E\left[(1+\sir_t)^{-\theta T}\right]$
\begin{eqnarray*}
	\Psi(\theta T) = 1 - \int_{0}^{1}\Ps(t^{-\frac{1}{\theta T}}-1)\d t.
\end{eqnarray*}

The distribution of the effective capacity can be calculated using Gil-Pelaez theorem, which involves numerical integration and does not provide much insight on the behavior of the effective capacity. For that, we establish below bounds on the distribution using classical concentration inequalities. The simplest upper bound on the complementary cdf (ccdf) follows from Markov's inequality:
\begin{eqnarray}
\P^{!0}(\mathcal{R}(\theta T) > x) \leq \frac{\E[\mathcal{R}(\theta T)]}{x}
\end{eqnarray}
where the first moment can be upper bounded as follows:
\begin{eqnarray*}
&&\hspace{-5mm}\E[\mathcal{R}(\theta T)] = -\frac{1}{\theta T} \E[\log \Psi(\theta T)] \stackrel{(a)}{\leq}  -\frac{1}{\theta} \log \E[\Psi(\theta T)] \\
&&\stackrel{(b)}{=} -\frac{1}{\theta T} \log \left(1 - \int_{0}^{1}\E^{!o}[\Ps(t^{-\frac{1}{\theta T}}-1)]\d t \right) \\
&&\stackrel{(c)}{=} -\frac{1}{\theta T} \log \left(1 - \int_{0}^{1}e^{-\lambda \mathcal{I}}\d t \right) \\
&&= -\frac{1}{\theta T} \log \left(1 - \int_{0}^{1}e^{-\lambda\pi p r^2\mathcal{C}(\alpha)(t^{-\frac{1}{\theta T}}-1)^{2/\alpha}} \d t \right) \\
&& = -\frac{1}{\theta T} \log \left(1 - \theta T \int_{0}^{\infty}\frac{e^{-\lambda\pi p r^2\mathcal{C}(\alpha)y^{2/\alpha}}}{(1+y)^{1+\theta T}} \d y \right) 
\end{eqnarray*}
where $\mathcal{I} = \displaystyle \int_{\mathbb{R}^2}\left(1 - \frac{1}{1+(t^{-\frac{1}{\theta T}}-1) r^\alpha\|x-R_o\|^{-\alpha}}\right) \d x$, and $\mathcal{C}(\alpha) = \frac{2}{\alpha}B(1-2/\alpha,2/\alpha)$ with $B(a,b)$ denoting the beta function. Step (a) follows using Jensen's inequality, (b) by exchanging expectation and integration order (Fubini-Tonelli's Theorem), (c) from the probability generating functional of the PPP. The first moment can be further bounded applying Cauchy-Schwarz inequality in the integral inside the logarithm. 

A lower bound on the ccdf of the effective capacity can be found using the Paley-Zygmund or the reverse Markov inequality, which however involves the calculation of the mean and the variance of $\mathcal{R}(\theta T)$. Using the upper bound on the Mellin transform based on the nearest interferer, we can establish the following lower bound on the ccdf:
\begin{eqnarray*}
\P^{!0}(\mathcal{R}(\theta T) > x)  & = & \P^{!0}(\setM_{\gamma}(1-\theta T) < e^{-x\theta T}) \\
& \geq & \P^{!0}(\mathcal{M}_{\gamma}^{u2}(1-\theta T) < e^{-x\theta T}) \\
& = & \P^{!0}\left(Z^{-1/2} < \zeta \right) = 1 - e^{-\lambda\pi r^2\zeta^{\frac{4}{\alpha}}}
\end{eqnarray*}
where $\zeta = \frac{e^{-x\theta T}-1}{\theta T(1+2\theta T)^{-1/2}}$.

\section{Numerical Results} \label{sec:num}
In this section, we validate the above analysis and provide numerical evaluation of the delay performance. The duration of a slot is set to $T=1$~ms and the blocklength is $N=100$. For the Poisson bipolar network we have a density of $\lambda = 1$ node/km$^2$ and pathloss exponent $\alpha = 3.5$. 

\begin{figure}[ht]
	\centering
	\includegraphics[width=0.85\columnwidth]{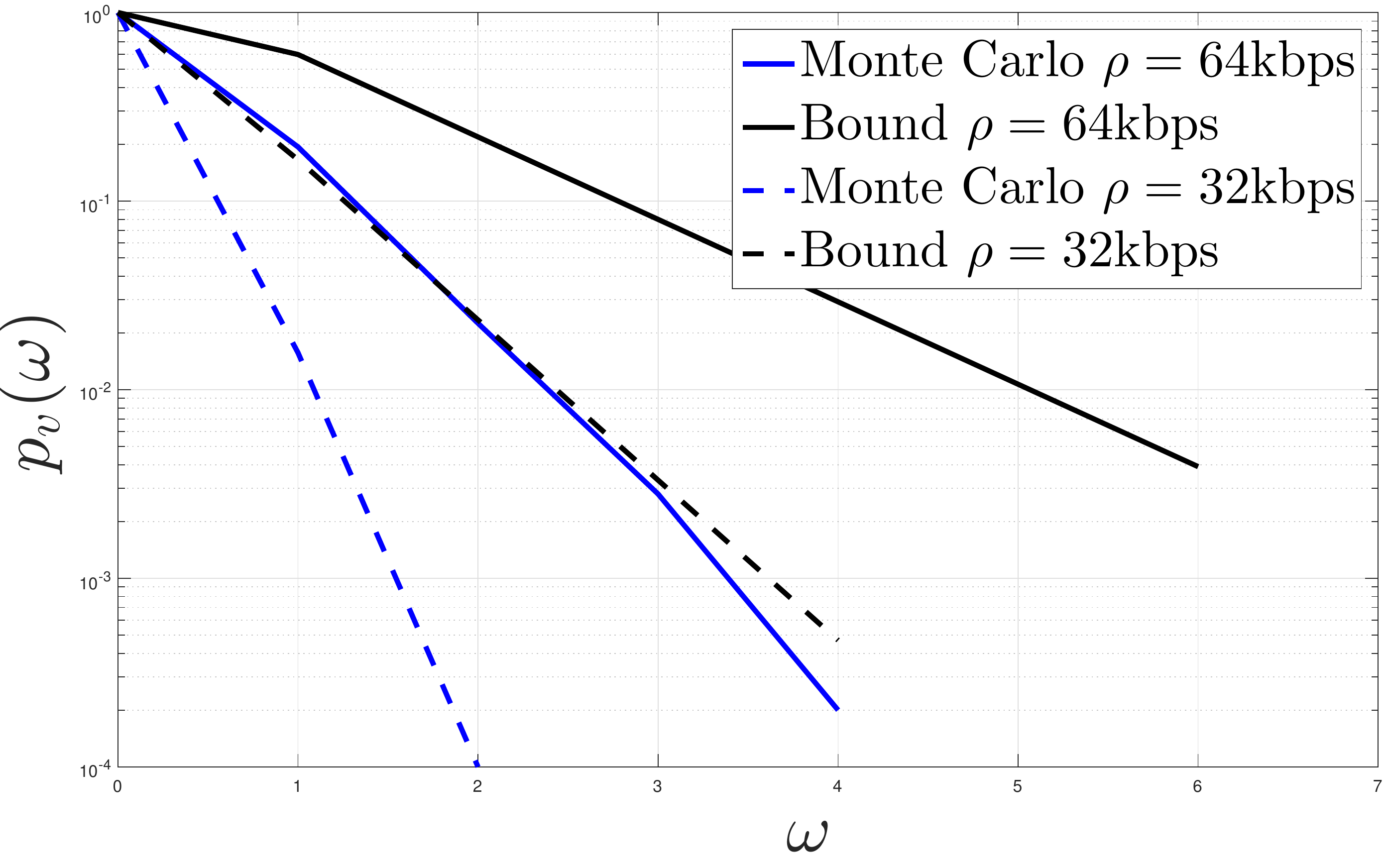}%
	\caption{Delay violation probability and associated bound as a function of the target delay for $\rho = 32$~kbps and $\rho = 64$~kbps.}
	\label{fig1}
\end{figure}

We start by validating the upper bound on the delay violation probability with Monte Carlo simulations. In Figure~\ref{fig1}, we compare the delay violation probability and its bound for link distance $r = 0.3$ km. We corroborate that the analytical bound follow the trend of the simulated curve, having a difference of about $1 - 2$~ms (equivalent to one to two slots).

\begin{figure}[ht]
	\centering
	\includegraphics[width=0.85\columnwidth]{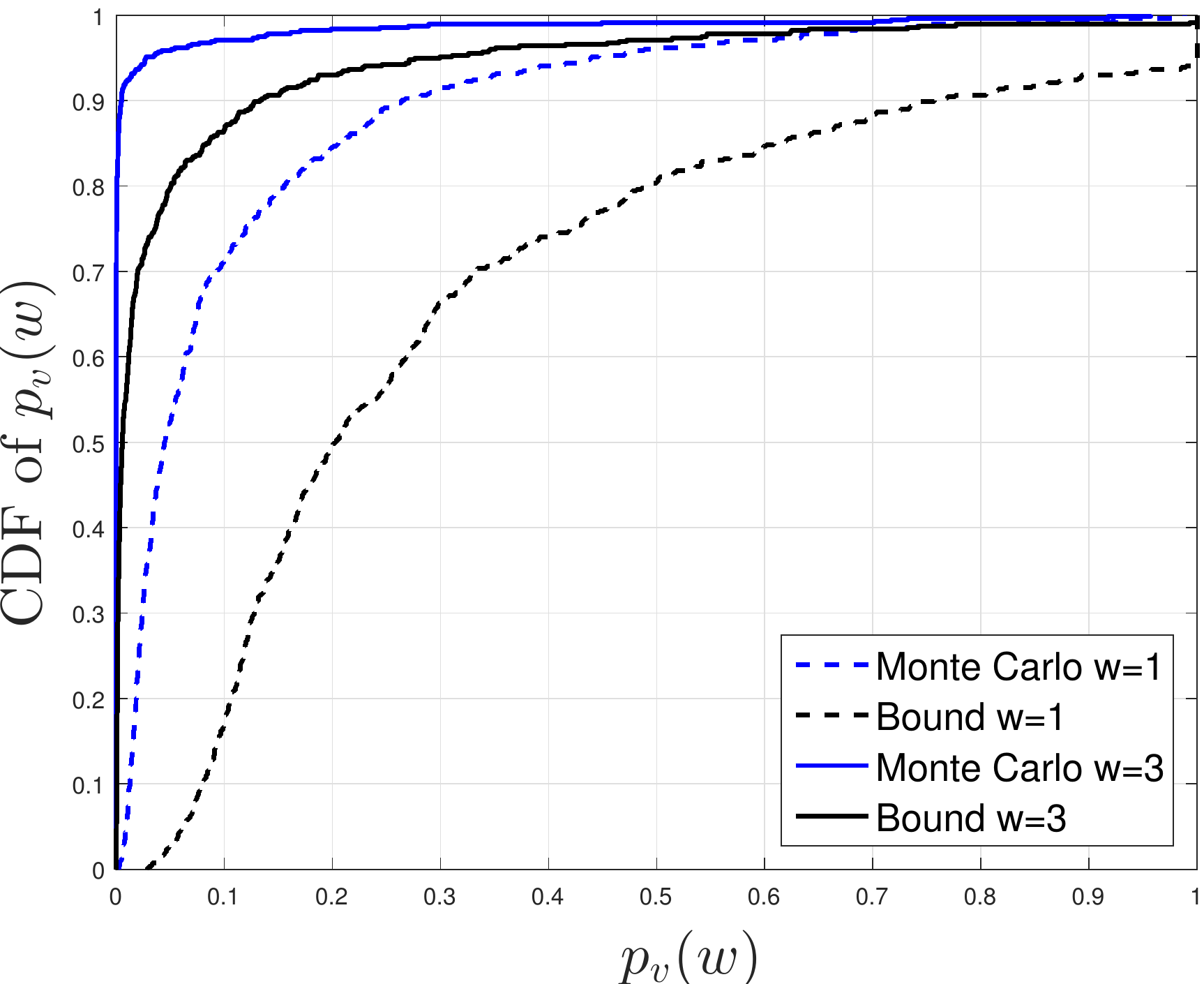}%
	\caption{Distribution of delay violation probability for $\omega = 1$ms and $\omega = 3$ms.}
	\label{fig2}
\end{figure}

In Figure~\ref{fig2}, we compare the distribution of the delay violation probability and that of the analytical bound for $\rho = 64$~kbps $r = 0.2$ km. We observe that the analytical bound becomes tighter for $\omega$ increasing.

\begin{figure}[ht]
	\centering
	\includegraphics[width=0.85\columnwidth]{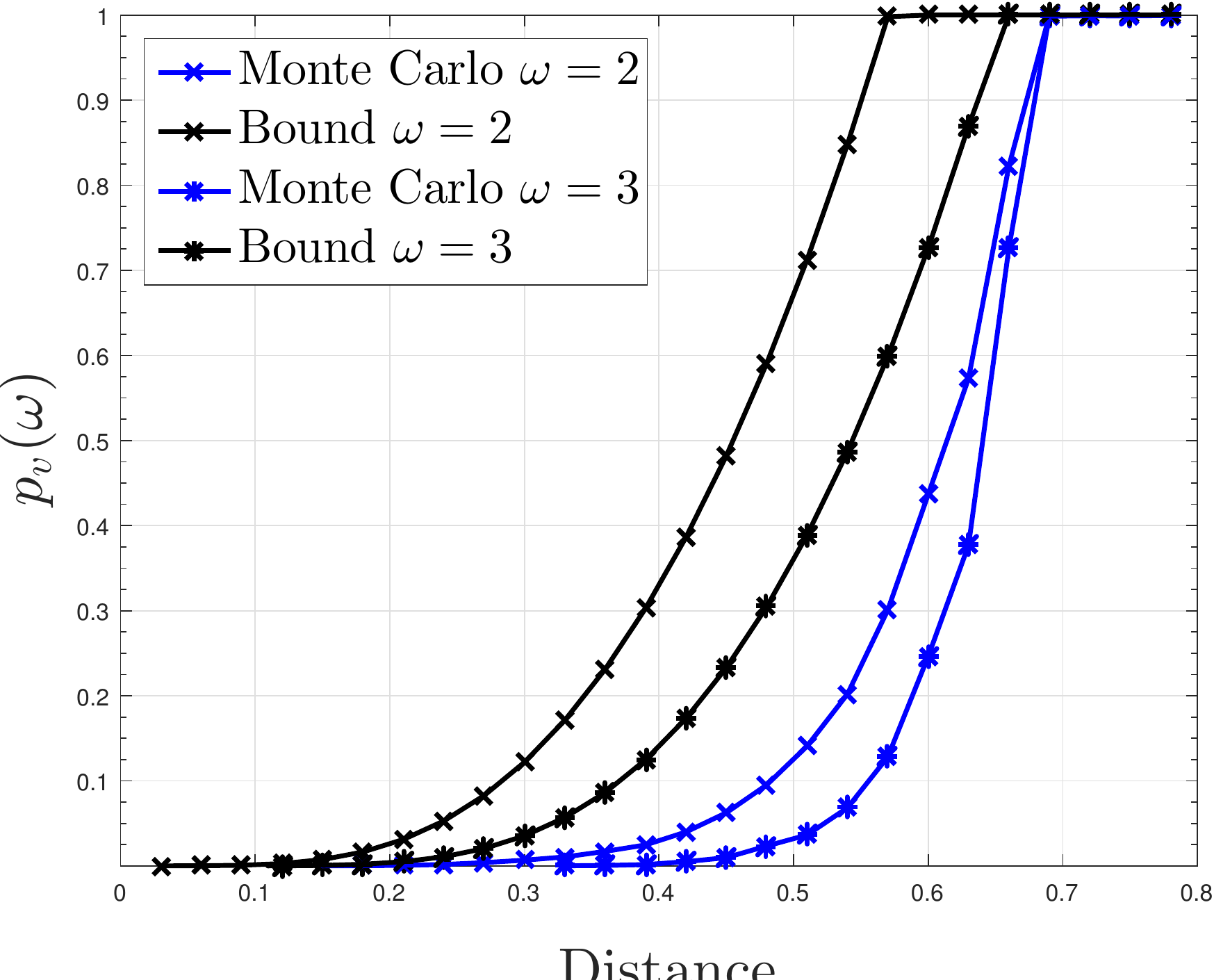}%
	\caption{Delay violation probability and associated bound as a function of the link distance $r$ for $\rho = 64$~kbps.}
	\label{fig3}
\end{figure}

Finally, in Figure~\ref{fig3}, we plot the delay violation probability and its analytical bound as a function of the inter-node distance $r$ for two different values of $\omega$. As expected, the more stringent the delay constraint is, the closer the transmitter and its intended receiver should be. Alternatively, for fixed link distance, tighter delay constraints can be guaranteed for lower density of interferers $\lambda$. 

\section{Conclusions} \label{sec:conc}
We have investigated the delay performance of large wireless networks in the presence of statistical QoS constraints. We have characterized the distribution of the conditional delay violation probability and effective capacity in Poisson bipolar networks. Our results provide useful insight into providing delay guarantees in random spatial networks. From a broader perspective, this paper is a first attempt to combine stochastic network calculus with stochastic geometry as a means to quantify the delay in wireless networks with spatial randomness. 

\section*{Acknowledgment}
The work of N. Pappas was supported in part by the Center for Industrial Information Technology (CENIIT).

\smallskip

\appendices
\section{Proof of Proposition~\ref{PropMT}} \label{sec:A1}
We start by deriving the conditional success probability $\Ps(\xi)$, i.e., the probability that a transmission will be successful by exceeding $\xi$ conditioned upon $\Phi$ \cite{MetaDistr}. 

\begin{eqnarray*}
\Ps(\xi) \hspace{-3mm} & = & \P(\sir_t >\xi\mid\Phi) \\
& = &\P\left(h_{oR_o}^{t}r^{-\alpha}>\xi I_{R_o,t}\mid\Phi\right) \\
&\stackrel{(a)}{=}&\E\left(\exp\left(-\xi r^\alpha I_{R_o,t}\right)\mid\Phi\right) 
\\
&\hspace{-5mm}=& \hspace{-6mm}\E\Big(\exp\Big(-\sum_{x\in\Phi\backslash\{o\}}\!\!\xi r^\alpha h_{xR_o}^{t}\|x-R_o\|^{-\alpha}\mathds{1}(x\in\Phi_t)\Big)\mid\Phi\Big) \\
&=&\hspace{-4mm} \prod_{x\in\Phi\backslash\{o\}}\!\!\!\!\E\left(\exp\left(-\xi r^\alpha h_{xR_o}^{t}\|x-R_o\|^{-\alpha}\mathds{1}(x\in\Phi_t)\right)\mid\Phi\right)\\
&\stackrel{(b)}{=}\!\!\!\!&\prod_{x\in\Phi\backslash\{o\}}\!\!\!\!\!\!\Big(\frac{p}{1+\xi r^\alpha\|x-R_o\|^{-\alpha}}+1-p\Big). \\
\end{eqnarray*}
where $(a)$ and $(b)$ follows because the fading coefficients are i.i.d. random variables with exponential distribution of unit mean.

Therefore, using integration by parts, we have 
\begin{eqnarray*}
	\mathcal{M}_{\gamma}(s) & = &	\int_0^{\infty}(1+y)^{s-1} \d \P(\sir < y \mid \Phi) \\
	& = &	-(s-1) \int_0^{\infty}(1+y)^{s-2} \P(\sir < y \mid \Phi) \d y \\
	& = & 1 +(s-1) \int_0^{\infty}(1+y)^{s-2}   \Ps(y) \d y.
\end{eqnarray*}

\addcontentsline{toc}{chapter}{References}
\bibliographystyle{IEEEtran}
\bibliography{IEEEabrv,ref_Huawei,bib_snc}

\end{document}